\documentclass[floats,floatfix,showpacs,amssymb,prd,twocolumn,superscriptaddress,nofootinbib,nolongbibliography,reprint]{revtex4-2}

\usepackage{amssymb, amsmath, nccmath, verbatim, mathtools, needspace, enumitem, etoolbox, graphicx, physics, microtype, afterpage, bigints, gensymb, tabularx, xspace, longtable, booktabs}

\usepackage[dvipsnames, usenames]{xcolor}
\definecolor{linkcolor}{rgb}{0.0,0.3,0.5}
\definecolor{dodgerblue}{HTML}{1E90FF}
\usepackage[unicode, colorlinks=true, linkcolor=linkcolor, citecolor=linkcolor, filecolor=linkcolor,urlcolor=linkcolor, pdfusetitle]{hyperref}
\usepackage[all]{hypcap}
\usepackage[T1]{fontenc}
\usepackage[utf8]{inputenc}
\usepackage{orcidlink}
\usepackage{natbib}
\usepackage{url}
\usepackage{cleveref, multirow}
\usepackage[caption=false]{subfig} 
\usepackage[normalem]{ulem} %
\usepackage{bibunits}

\newcommand{\ssim}{\mathchar"5218\relax\,}

\makeatletter
\newcommand*{\balancecolsandclearpage}{\close@column@grid \cleardoublepage \twocolumngrid}
\makeatother

\allowdisplaybreaks 

\newcommand{\milan}{\affiliation{Dipartimento di Fisica ``G. Occhialini'', Universit\'a degli Studi di Milano-Bicocca, Piazza della Scienza 3, 20126 Milano, Italy}}
\newcommand{\infn}{\affiliation{INFN, Sezione di Milano-Bicocca, Piazza della Scienza 3, 20126 Milano, Italy}}

\newcommand{\ecat}{\ensuremath{e_{\rm c}}\xspace}
\newcommand{\egw}{\ensuremath{e_{\rm gw}}\xspace}

\begin{document}

\title{
Orbital eccentricity in general relativity from catastrophe theory
}

\author{Matteo Boschini$\,$\orcidlink{0009-0002-5682-1871}}
\email{m.boschini1@campus.unimib.it}

\milan \infn

\author{Nicholas Loutrel$\,$\orcidlink{0000-0002-1597-3281}}

\milan \infn 

\author{Davide Gerosa$\,$\orcidlink{0000-0002-0933-3579}}

\milan \infn

\author{Giulia Fumagalli$\,$\orcidlink{0009-0004-2044-989X}}

\milan \infn


\date{\today}

\begin{abstract}
While the orbital eccentricity is a key feature of the gravitational two-body problem, providing an unambiguous definition in general relativity poses significant challenges. Despite such foundational issue, the eccentricity of binary black holes has important implications %
in gravitational-wave astronomy. We present a novel approach to consistently define the orbital eccentricity in general relativity, grounded in the mathematical field of catastrophe theory. Specifically, we identify the presence of catastrophes, i.e., breakdowns of the stationary-phase approximation, in numerical relativity waveforms and exploit them to develop a robust and fully gauge-invariant estimator of the eccentricity. Our procedure does not require orbital fitting and naturally satisfies the Newtonian limit. %
The proposed eccentricity estimator agrees with and generalizes a previous proposal, though with a fully independent derivation.
We extract gauge-free eccentricity estimates from about 100 numerical-relativity simulations and find that the resulting values are systematically lower compared to those reported alongside the simulations themselves. 
\end{abstract}

\maketitle

\section{Introduction} 
Generic solutions to the two-body problem in Newtonian gravity are represented by conic sections, where the orbital eccentricity $e$ discriminates between closed and unbound orbits. %
Conversely, defining the eccentricity in general relativity (GR) is a highly nontrivial problem: radiation reaction and periastron precession both lead to nonclosed orbits and gauge effects are hard to eradicate~\cite{2019CQGra..36b5004L,Blanchet:2013haa,2023PhRvD.108j4007S}.  %
 
Achieving a consistent definition of the orbital eccentricity is a pressing issue for both current and future gravitational-wave (GW) observations. For stellar-mass black-hole (BH) binaries presently targeted by LIGO/Virgo~\cite{2019PhRvX...9c1040A,2021PhRvX..11b1053A,2024PhRvD.109b2001A,2023PhRvX..13d1039A}, and in the future by third-generation detectors~\cite{2020JCAP...03..050M,2019BAAS...51g..35R}, the orbital eccentricity is a telltale sign of dynamical formation in dense environments such as clusters or disks~\cite{2021hgwa.bookE..16M, 2022PhR...955....1M}. For supermassive BHs targeted by LISA~\cite{2024arXiv240207571C}, eccentricity is an indicator for gas accretion and recent three-body interactions~\cite{2021hgwa.bookE..18B,2021NatRP...3..732V}.
A BH binary detection with eccentricity that is confidently constrained away from zero is arguably one of the most long-awaited GW events, and there is already some tentative evidence in current data~\citep{2022ApJ...940..171R,2024arXiv240414286G,2024ApJ...972...65I}. At the same time, neglecting eccentricity when present can deeply impact both the astrophysics \cite{2024arXiv240514945F}
 and testing GR~\cite{2024PhRvD.109h4056S,2024arXiv240502197G} science cases of GW detectors.

Such a large scientific payoff is prompting major efforts to include eccentricity in waveform approximants ---a task that remains challenging in particular when combined with spin precession \cite{2016PhRvD..93f4031T, 2018CQGra..35n4003V, 2018CQGra..35w5006M, 2019CQGra..36r5003M, 2020CQGra..37g5008L, 2021PhRvD.103f4013N, 2021PhRvD.103f4022I, 2021PhRvD.103j4014H, 2021PhRvD.104f4047K, 2022CQGra..39c5009L, 2022PhRvD.105d4035R, 2022PhRvD.105f4010C, 2022CQGra..39n5004L,
2024PhRvD.110h4001N,
2024PhRvD.110d4044A,
2024PhRvD.110b4031G,
2024PhRvD.109f4030D, 2024CQGra..41v5002L, 2024CQGra..41s5019L,
2024arXiv240913866P}. In numerical relativity (NR),  several of the public 
catalogs now contain simulations of eccentric
 binary BHs \cite{2019PhRvD.100f4003H, 2019CQGra..36s5006B, 2022PhRvD.105l4010H, 2022PhRvD.106l4040R, 2023arXiv230900262F, 2024arXiv240918728F}, although with a somewhat sparse coverage of the parameter space, especially at high eccentricities.  
Currently, each NR effort has implemented a different definition of eccentricity. For instance, (i) the Simulating eXtreme Spacetimes (SXS) catalog~\cite{2019CQGra..36s5006B,Mroue:2013xna} reports values of $e$ that rely on a post-Newtonian (PN) fit of the orbital-frequency time derivative over the first $\ssim 2$ orbits past junk radiation \cite{2011PhRvD..83j4034B, 2019CQGra..36s5006B}, (ii) the Rochester Institute of Technology (RIT) group \cite{2022PhRvD.105l4010H, 2024arXiv240918728F} employs a definition $e \propto r^2 \dd^2 r/ \dd t^2$  which uses the coordinate separation $r$ %
and (iii) the MAYA catalog \cite{2023arXiv230900262F} defines $e$ through the orbital frequency as $e = (\Omega - \Omega_{c})/(2\Omega_{c})$, where $\Omega = d\phi/dt$ with $\phi$ as the azimuthal coordinate and $\Omega_{c}$ as the orbital frequency of a quasicircular simulation~\cite{2019PhRvD..99b3003R}. All these definitions are gauge dependent due to reliance on coordinate quantities.  
Recent work by \citeauthor{2023PhRvD.108j4007S}~\cite{2023PhRvD.108j4007S} (see also Ref.~\cite{2022PhRvD.106l4040R}) %
provides a definition that uses the (2,2)-mode waveform frequency extracted at future null infinity. While this moves the signpost toward a gauge-independent notion of the orbital eccentricity, it still requires a manual adjustment to obtain the correct Newtonian limit.

In this paper, we present a novel approach to consistently define the binary eccentricity in GR, which has solid foundation in the branch of mathematics called ``catastrophe theory''~\cite{DBLP:books/daglib/0070571, 5d38e71d-3dd9-3f44-bf01-2ae08cb112f4, Saunders_1980}. Part of the more general singularity theory, catastrophe theory deals with functions depending on both variables and parameters, characterizing how continuous and smooth variations of the latter can cause abrupt transitions of the former. These discontinuities, or ``catastrophes,'' are structural critical points in the function domain where the first derivative and one or more higher-order derivatives %
with respect to the variables
vanish. %
Catastrophe theory has been applied to systems ranging from physics to biology~%
\cite{0a7b6414-4db8-3303-8632-1d9bc8fc3f79, ROOPNARINE2008531,1981PhyD....2..245S,2002PhR...356..229A}; %
in GW astronomy, a toy application was first presented by one of us \cite{2023CQGra..40u5004L}. For the dominant emission mode, our rigorous derivation reduces to the expression of Ref.~\cite{2023PhRvD.108j4007S}, which is very encouraging. At the same time, with the methodology presented here, it is straightforward to extract the eccentricity from other subdominant contributions to the emitted gravitational radiation and average the result using the relative contribution to the emitted energy. 

Our paper is organized as follows. In Sec.~\ref{sec:two}, we derive the proposed definition, applying the catastrophe formalism to PN analytic waveforms.  In Sec.~\ref{sec:three}, we present the first evidence that catastrophes are present in NR waveforms and show how these catastrophes can be exploited to obtain a solid estimator of the orbital eccentricity. %
We draw some concluding remarks in Sec.~\ref{sec:six}. 
We use geometric unit where $G=c=1$.

\section{Post-Newtonian waveforms}
\label{sec:two}

Let us first apply our catastrophe analysis to analytic PN waveforms where the eccentricity can be confidently defined \cite{1985AIHPA..43..107D}. This is useful to both gain intuition and identify the correct mapping between frequency and eccentricity that can then be applied to NR. 
\subsection{Catastrophes in post-Newtonian theory}

Catastrophes in the waveforms produced by eccentric binaries were first considered in Ref.~\cite{2023CQGra..40u5004L} by decomposing the two GW polarizations into their orbital harmonic modes and Fourier-transforming each mode individually. This is appropriate for PN waveforms but does not generalize to NR.  We instead consider the complex strain %
and factor out the angular dependence on the inclination angle $\iota$ and the polarization angle $\upsilon$ using spin-weighted spherical harmonics,
\begin{equation}
\label{eq:sp_harm}
	h(t) = h_+(t) - i h_{\times}(t) = \sum_{l=2}^{\infty}\sum_{m=-l}^{l} {_{-2}}Y_{lm}(\iota,\upsilon) h_{lm}(t)\,.
\end{equation}
We focus on three GW harmonics at the corresponding leading PN orders: $(l,m)= {\rm (2,2)^{0PN}}$, ${\rm (3,3)^{0.5PN}}$, and ${\rm (4,4)^{1PN}}$. Note that the procedure below does not depend on the sign of $m$.
Defining $h_{lm,+} = {\rm Re}\;h_{lm}$ and $h_{lm,\times} = {\rm Im}\;h_{lm}$, the relevant expressions for the GW polarizations are given by \cite{Poisson_Will_2014} %
\begin{widetext}
	\begin{align}
		\label{eq:h_22+}
		h^{\rm0PN}_{22,+} = & -\frac{2 \sqrt{\frac{\pi}{5}}   M^2}{d\,a(1-e^2)} \frac{q}{(1+q)^2}\left[4\cos2\varphi + e\left(5\cos\varphi + \cos3\varphi\right) + 2e^2\right]\, , \\
		\label{eq:h_22x}
		h^{\rm0PN}_{22, \times}  = & \frac{4 \sqrt{\frac{\pi }{5}}   M^2 }{d\,a(1-e^2)} \frac{q}{(1+q)^2} \sin\varphi \left[4\cos\varphi +e \left(3 + \cos2\varphi\right) \right] \, ,
\\
		\label{eq:h_33+}
		h^{\rm0.5PN}_{33,+}  = & \frac{\sqrt{\frac{\pi }{42}}  M^{5/2}}{d \left[a{\left(1- e^2\right)}\right]^{3/2}}
		\frac{q(1-q)}{(1+q)^3} 
		\sin\varphi  \left[18 + 36\cos2\varphi + e(80\cos\varphi + 20\cos3\varphi)+e^2(26 + 17\cos2\varphi+3\cos4\varphi)\right]\, , \\
		\label{eq:h_33x}
		h^{\rm0.5PN}_{33, \times}  = & \frac{\sqrt{\frac{\pi }{42}}  M^{5/2}}{d \left[a{\left(1- e^2\right)}\right]^{3/2}} 
		\frac{q(1-q)}{(1+q)^3} 
		 \left[36\cos3\varphi + e(60\cos2\varphi + 20\cos4\varphi) + e^2(35\cos\varphi + 14\cos3\varphi + \cos5\varphi) + 8e^3\right]\, ,
\\
		\label{eq:h_44+}
		h^{\rm1PN}_{44,+}  = & \frac{\sqrt{\frac{\pi }{7}}  M^3 }{72 d \, a^2 \left(1-e^2\right)^2}
		\frac{q(1-q+q^2)}{(1+q)^4}
		 \bigl[512\cos4\varphi + e(1076\cos3\varphi + 452\cos5\varphi)  + e^2(860\cos2\varphi+ 568\cos4\varphi\,  + \notag \\
		& 
		+ 140\cos6\varphi) + e^3(315\cos\varphi + 189\cos3\varphi + 81\cos5\varphi + 15\cos7\varphi )+ 48e^4\bigr]\, , \\
		\label{eq:h_44x}
		h^{\rm1PN}_{44, \times}  = & -\frac{\sqrt{\frac{\pi }{7}} q M^3 }{36 d\,a^2 \left(1-e^2\right)^2 }%
				\frac{q(1-q+q^2)}{(1+q)^4}
		 \sin\varphi \bigl[512(\cos\varphi + \cos3\varphi) + e(764 + 1528\cos2\varphi + 452e\cos4\varphi) \, + \notag \\
		& + e^2(1568\cos\varphi + 708\cos3\varphi +140\cos5\varphi) + e^3(300+285\cos2\varphi+96\cos4\varphi+15\cos6\varphi) \bigr]\, ,
	\end{align}
\end{widetext}
where $M$ is the total mass, $q\leq 1$ %
is the mass ratio, %
$a$ is the semimajor axis, $e$ is the eccentricity, $d$ is the luminosity distance to the source, and $\varphi$ is the orbital phase. %
For each of these modes, we %
express the complex strain as %
\begin{align}
	h_{lm}(t) &= A_{lm}(t)e^{i\phi_{lm}(t)}\, ,
	\\
	A_{lm}(t) &= \sqrt{\left(h_{lm,+}\right)^{2} + \left(h_{lm,\times}\right)^{2}}\,,
	\\
	\phi_{lm}(t) &= \tan^{-1}\left(\frac{h_{lm,\times}}{h_{lm,+}} \right)\,,
\end{align} 
and construct the Fourier phase as 
\begin{equation}\Psi(t) = 2\pi{f}t - \phi_{lm}(t)\,.
\end{equation}

The general approach to solving the binary dynamics relies on the stationary-phase approximation (SPA) where one identifies $t_{\star}$ such that $\dot{\Psi}(t_{\star}) = 0$.  It turns out that, for eccentric binaries, stationary points exist, and thus the SPA holds, only for frequencies $f$ in a bound interval $f^-\leq f\leq f^+$.
In the limiting case where $f = f^\pm$, one has $\ddot{\Psi}(t_{\star}) = 0$, which identifies a catastrophe.

To this end, we %
first promote the semimajor axis $a$, the eccentricity $e$, and the orbital phase $\varphi$ to functions of time, where the derivatives $\dd a / \dd t$ and $\dd e /\dd t$ are evaluated according to the equations by \citeauthor{1964PhRv..136.1224P}~\cite{1964PhRv..136.1224P}, while $\dd {\varphi} / \dd t$  is given by Kepler's law, specifically 
\begin{align}
	\label{eq:dadt}
	\frac{\dd a}{\dd t} &= -\frac{64}{5} \frac{M^3}{a^3} \frac{q}{(1+q)^2} \frac{\left(1 + \frac{73}{24} e^{2} + \frac{37}{96}e^{4}\right)}{(1-e^{2})^{7/2}} \, ,
	\\
	\label{eq:dedt}
	\frac{\dd e}{\dd t} &= -\frac{304}{15}  \frac{M^3}{a^4} \frac{q}{(1+q)^2}  \frac{e\left(1 + \frac{121}{304} e^{2}\right)}{(1-e^{2})^{5/2}}\, ,
	\\
	\label{eq:dvarphidt}
	\frac{\dd \varphi}{\dd t} &= \frac{M^{1/2}}{a^{3/2}} \frac{\left(1 + e \cos\varphi\right)^{2}}{(1-e^{2})^{3/2}}\, .
\end{align}

The frequencies at the stationary times, which we denote $f_{lm}$, are given by

\begin{figure}
  		\includegraphics[width=\columnwidth] {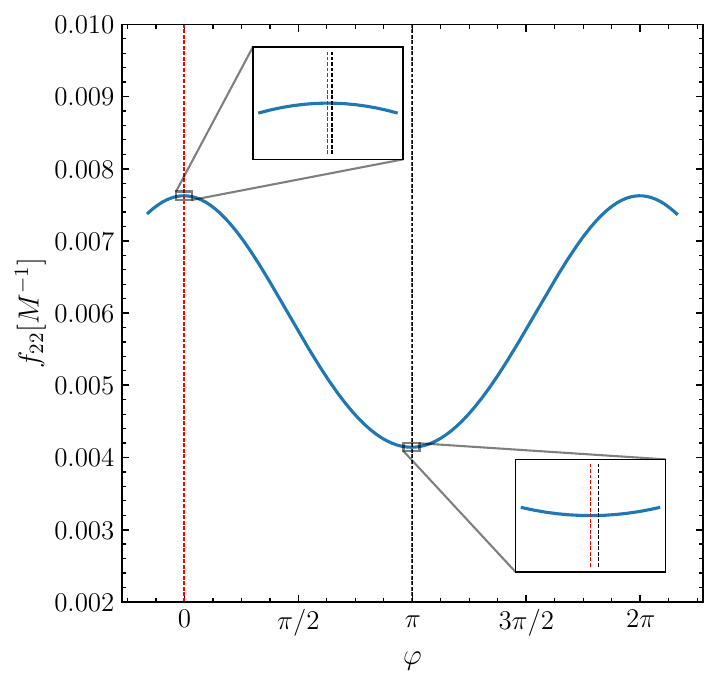} 
  		\caption{Stationary frequencies of the (2,2) mode in the PN approximation. The blue curve shows the evolution of $f_{22}$ from Eq.~(\ref{eq:f_22star}) as a function of the orbital phase $\varphi$ for a system with separation $a=15M$,  eccentricity $e=0.2$,  
		and mass ratio $q=1$. 
The two black dashed lines identify the pericenter and apocenter passage $\varphi=(0,\pi)$. The two red dashed lines indicate the extrema of $f_{22}$. The separation between the two sets of lines   is a 2.5PN correction (see insets).}
  		\label{fig:one}
		\end{figure}

\begin{widetext}
	\begin{align}
				\label{eq:f_22star}
		f_{22}^{\rm 0PN} &= \frac{\sqrt{M}}{\pi \left[a \left(1-e^2\right)\right]^{3/2}}  \Biggl\{ \left(1 + e\cos\varphi\right)^3 \left[8 + 12e\cos\varphi + e^2(1 + 3\cos 2\varphi) \right]- \frac{M^{5/2}}{15  a^{5/2}}\frac{q}{(1+q)^2} \frac{e\left(304 + 121e^2\right) }{1-e^2}  \sin\varphi 
		\notag\\ 
		& \times \left[4 + 8e\cos\varphi + e^2(3 + \cos 2\varphi)  \right]\Biggr\} \Big[ 8 + 24e\cos\varphi + 13e^2(1 + \cos 2\varphi) + 2e^3(5\cos\varphi + \cos 3\varphi) + 2e^4 \Big]^{-1} \\
				\label{eq:f_33star}
		f_{33}^{\rm 0.5PN} &= \frac{\sqrt{M}}{\pi \left[a \left(1-e^2\right)\right]^{3/2}}  \Biggl\{ \left(1 + e\cos\varphi\right)^3 \Bigl[972 + 2988e\cos\varphi + 4e^2\left(373 + 432\cos2\varphi\right)+ e^3\left(897\cos\varphi + 515\cos3\varphi \right) 
		 \notag\\
		&+ 4e^4\left(4 + 27\cos2\varphi + 15\cos4\varphi \right) \Bigr]- \frac{4M^{5/2}}{15  a^{5/2}}\frac{q}{(1+q)^2} \frac{e\left(304 + 121e^2\right) }{1-e^2}  \sin\varphi \Bigl[90 + 288e\cos\varphi \notag\\
		& + e^2\left(179 + 173\cos2\varphi\right)
		+ 40e^3\left(4\cos\varphi + \cos3\varphi\right) +e^4\left(26 + 17\cos2\varphi + 3\cos4\varphi\right)\Bigr]\Biggr\} 
		\notag \\
		 &\times \Big[ 648 + 1880e\cos\varphi +8e^2\left(313 + 321\cos2\varphi\right) +16e^3\left(205\cos\varphi +73\cos3\varphi \right)  \notag\\
		&+ e^4\left(715 + 1012\cos2\varphi + 265\cos4\varphi\right) + 8e^5\left(35\cos\varphi + 14\cos3\varphi + 3\cos5\varphi\right) + 32 e^6 \Big]^{-1}\\
		\label{eq:f_44star}
		f_{44}^{\rm 1PN} &= \frac{4\sqrt{M}}{\pi \left[a \left(1-e^2\right)\right]^{3/2}}  \Biggl\{ \left(1 + e\cos\varphi\right)^3 \Bigl[65536 + 305664e\cos\varphi + 24e^2\left(11396 + 12511\cos2\varphi \right) + 4e^3(96627\cos\varphi 
		\notag\\
		&+ 40087\cos3\varphi ) + 6e^4\left(14391 + 22594\cos2\varphi + 8275\cos4\varphi \right)+ 3e^5\left(9936\cos\varphi + 8069\cos3\varphi + 2835\cos5\varphi \right) 
		 \notag\\
		&+ 18e^6\left(8 + 89\cos2\varphi + 100\cos4\varphi + 35\cos6\varphi \right)\Bigr]- \frac{8M^{5/2}}{5 a^{5/2}}\frac{q}{(1+q)^2} \frac{e\left(304 + 121e^2\right) }{1-e^2}  \sin\varphi \Bigl[3328 + 15360e\cos\varphi 
		 \notag\\
		&+ 8e^2\left(1787 + 1820\cos2\varphi \right)+ 4e^3\left(5269\cos\varphi + 1827\cos3\varphi \right) +2e^4\left(2903 + 3942\cos2\varphi + 1003\cos4\varphi \right) \notag\\
		&+ 8e^5\left(392\cos\varphi + 177\cos3\varphi + 35\cos5\varphi \right) + 3e^6\left(100 + 95\cos2\varphi + 32\cos4\varphi + 5\cos6\varphi \right)\Bigr]\Biggr\} \notag \\
		&\times \Big[ 131072  + 782336e\cos\varphi + 144e^2\left(6749 + 6933\cos2\varphi\right) + 32e^3\left(62337\cos\varphi + 22135\cos3\varphi \right) + 8e^4(97611 \notag\\
		&+ 135788\cos2\varphi + 37937\cos4\varphi ) + 24e^5\left(25010\cos\varphi + 13967\cos3\varphi + 3279\cos5\varphi \right) + 3e^6 (23622 + 39133\cos2\varphi 
		\notag\\
		&+ 18538\cos4\varphi + 3815\cos6\varphi ) + 144e^7\left(105\cos\varphi + 63\cos3\varphi + 27\cos5\varphi + 5\cos7\varphi \right) + 1152e^8\Big]^{-1}
	\end{align}
\end{widetext}
where $a$, $\varphi$, and $e$ are evaluated at $t_{\star}$. The leading-order terms $\propto a^{-3/2}$ arise from $\dd\varphi/\dd t$ and are thus of Newtonian order, while the next-to-leading-order terms  $\propto a^{-4}$ come from the adiabatic evolution of $(a,e)$ %
and are suppressed by 2.5PN orders. 
By computing $\ddot{\Psi}(t_{\star})$, it is straightforward to check that $f_{lm}$ reaches a maximum/minimum at $\varphi=(0,\pi) + {\cal{O}}\!\left(M/a\right)^{5/2}$, which are the apocenter and the pericenter passages, respectively. We dub those frequencies $f^{\pm}_{lm}$, which are explicitly given by 
\begin{align}
	\label{eq:f22_+-}
	 f_{22}^{\pm} &= \frac{2\sqrt{M}}{\pi a^{3/2}} \frac{\sqrt{1\pm e}}{(2\pm e) (1 \mp e)^{3/2}} + {\cal{O}}\left[\left(\frac{M}{a} \right)^{5}\right]\,, \\
	\label{eq:f33_+-}
	 f_{33}^{\pm} &= \frac{\sqrt{M}}{2\pi a^{3/2}} \frac{27\pm23e}{(1\!\mp \!e)(9\!\pm\!2e)\sqrt{1\! - \!e^2}} + {\cal{O}}\left[\left(\frac{M}{a} \right)^{5}\right]\,, \\
	\label{eq:f44_+-}
	 f_{44}^{\pm} &=\! \frac{\sqrt{M}}{\pi a^{3/2}} \frac{(1\pm e)(128\pm87e)}{(1\!\mp\! e)(64\!\pm\!63e\!+\!6e^2)\sqrt{1 \!-\! e^2}} 
	+{\cal{O}}\left[\left(\frac{M}{a} \right)^{5}\right]\!.
\end{align}
An example is illustrated in Fig.~\ref{fig:one}. Since the frequencies $f^{\pm}_{lm}$ correspond to $\ddot{\Psi}(t_{\star})=0$, these signal a breakdown of the SPA and mark the presence of a catastrophe.
 
The ratio
\begin{equation}
	\label{eq:ratio}
	\xi_{lm}(e) \equiv \frac{f_{lm}^+-f_{lm}^-}{f_{lm}^++f_{lm}^-}
\end{equation}
is a function solely of the eccentricity. %
From the equations above, one has
 \begin{align}
	\label{eq:r_22}
	& \xi_{22} = \frac{1}{2}e(3-e^2) + {\cal{O}}\left[\left(\frac{M}{a}\right)^{5}\right]\, , \\
	\label{eq:r_33}
	& \xi_{33} = \frac{396e - 46e^3}{243+107e^2} + {\cal{O}}\left[\left(\frac{M}{a}\right)^{5}\right]\, , \\
	\label{eq:r_44}
	& \xi_{44} = \frac{2e\left(6944-5700e^2+261e^4\right)}{8192-1513e^2-3669e^4} + {\cal{O}}\left[\left(\frac{M}{a}\right)^{5}\right]\, .
\end{align}
Note that one could arrive at the above relationship between $f^{\pm}$ and $e$ simply by using solely Newtonian physics, as is evidenced by the remaider being of a 5PN order. While the details of the catastrophe analysis predict the saddle points to be shifted away from pericenter/apocenter by 2.5PN order corrections, i.e. $\varphi = (0,\pi) + {\cal{O}}(M/a)^{5/2}$ (see Fig.~\ref{fig:one}), the mathematical structure of Eqs.~\eqref{eq:f_22star}--\eqref{eq:f_44star} conspires to force the shifts to a higher PN order in Eqs.~\eqref{eq:r_22}--\eqref{eq:r_44}. This should not be surprising, since the adiabatic radiation reaction introduces a second, longer timescale into the two-body problem, while eccentricity is an orbital quantity and, thus, defined on the shorter orbital timescale.

We now invert Eqs.~(\ref{eq:r_22})--(\ref{eq:r_44}) and select the only real root in $[0,1)$. For the (2,2) harmonic, we find:
\begin{align}
	\label{eq:e22}
	e_{22} &= \cos\left(\Theta\right) - \sqrt{3} \sin\left(\Theta\right)\, ,
	\\
	\Theta &= \frac{1}{3}\arctan\left(\frac{\sqrt{1-\xi_{22}^{2}}}{\xi_{22}}\right)\, .
\label{eq:theta22}
\end{align}
The analogous result for $(l,m)= (3,3)$ is
\begin{align}
	e_{33} &=  \frac{1}{138} \Bigl[ -107\xi_{33} + \sqrt{54648 + 11449\xi_{33}}\cos\left(\Phi \right) \notag \\
	                  & -\sqrt{163944 + 34347\xi_{33}}\sin\left(\Phi\right) \Bigr] \, ,
	              \label{eq:e33}
	\\
		\Phi &= \frac{1}{3}\arctan\Biggl(\frac{1242\sqrt{3}}{\xi_{33}}\, \notag \\
	&  \times \frac{\sqrt{35266176 - 31184003\xi_{33}^2 - 3675129 \xi_{33}^4}}{15712542 + 1225043 \xi_{33}^2}\Biggr)\, .
\label{eq:Phi33}
\end{align}
Finally, the expression for $e_{44}$ cannot be written down analytically because it involves solving a fifth-degree polynomial. However, it can easily be computed numerically.

The procedure we just highlighted maps the breaking points of the SPA $f_{lm}^{\pm}$ to the eccentricity $e$, where, crucially, the former can be extracted from waveform data at future null infinity and are thus gauge invariant. Our proposal is thus to promote the expressions above as a \emph{definition} of orbital eccentricity in GR. Hereafter, we refer to these eccentricity estimates as \ecat, where the subscript stands for ``catastrophe.'' %

\subsection{Orbit average}

\begin{figure}
 		\includegraphics[width=\columnwidth] {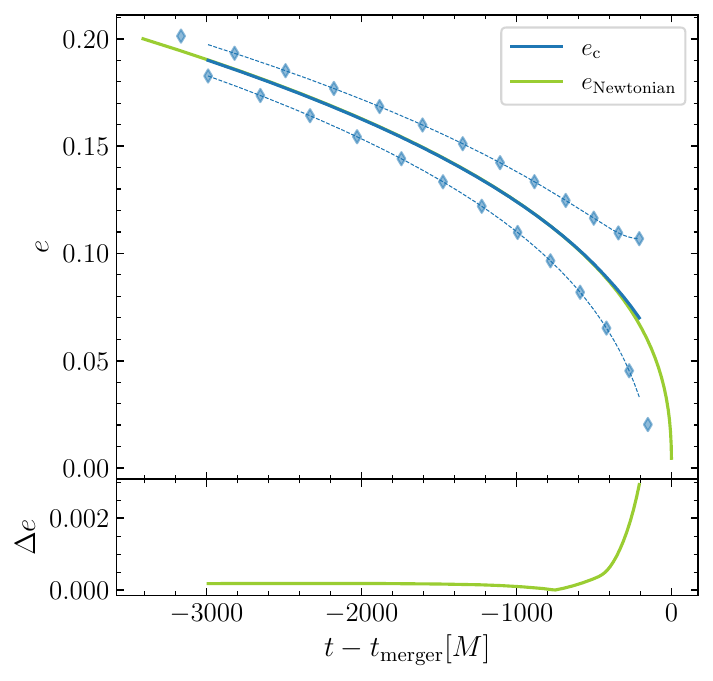} 
 		\caption{Eccentricity evolution for an analytic quadrupolar waveform with mass ratio $q=1$,  initial eccentricity $e=0.20$, and initial separation $a=15M$. This is a controlled experiment: the top panel shows the eccentricity \ecat  (blue curve) predicted by our catastrophe against the assumed evolution $e_{\rm Newtonian}$. %
	(green curve). The former is obtained by averaging estimates from two halves of each orbit (blue diamonds and dashed curves). The bottom panel shows the absolute difference between the two predictions. }	
		\label{fig:two}
\end{figure}

For internal consistency, we apply our catastrophe analysis to the (2,2) waveforms of Eqs.~(\ref{eq:h_22+}) and (\ref{eq:h_22x}), evolved accordingly to Eqs.~(\ref{eq:dadt})--(\ref{eq:dvarphidt}).  As an example, we select an equal-mass BH binary with initial conditions $a=15M$ and $e=0.20$.
Figure~\ref{fig:two} shows the reconstructed evolution of \ecat. In this controlled experiment, one must find $\ecat=e_{\rm Newtonian}$.  

From the procedure above, estimating the eccentricity requires a value of $f^+$ and a value of $f^-$---it can thus be performed every half orbit. One has that $f^+$ for the first half of the orbit is equal to $f^+$ for the second half of the orbit, but the former has a value of $f^-$ that is not equal to the value of $f^-$ of the latter. 
The reason is he radiation reaction, and it simply corresponds to one's intuition that the frequency increases as the binary inspirals toward merger. 

 The eccentricity estimates obtained using the evolution from apocenter to pericenter are higher than those computed using the evolution from pericenter to apocenter. We interpret these as upper and lower bounds, respectively, and assign the binary an eccentricity \ecat equal to their mean. This is not an arbitrary choice but is required for internal consistency with the Newtonian limit. 
Indeed, the average of these two estimates is in excellent agreement with the assumed evolution up to numerical errors of $\mathcal{O}\!\left(10^{-4}\right)$, at least up to the last few orbits before merger (amplitude maxima and minima cannot be reliably identified in that regime). A comparison in the early inspiral ($a=50M$) shows an even better agreement, with errors of $\mathcal{O}\!\left(10^{-7}\right)$.%

\section{Numerical relativity waveforms} 
\label{sec:three}

The procedure we just highlighted provides a practical recipe to extract the orbital eccentricity from a given gravitational waveform.  Let us now look for catastrophes in full GR using NR waveforms.

\subsection{Catastrophes in numerical relativity}

As an example, we consider the eccentric BH-binary simulation SXS:BBH:1372~\cite{2019CQGra..36s5006B} from the SXS catalog. The simulation evolves a nonspinning, unequal-mass,  eccentric binary for $t\sim4280M$, which corresponds to $\sim17.6$ orbits before merger. The following analysis uses the corresponding waveform extracted at the outermost radius with the highest available resolution. 

We select the three harmonics $(l,m)=(2,2)$, $(3,3)$, and $(4,4)$, and decompose the $+,\times$ polarizations into an amplitude  $A_{lm}(t)$ and a phase $\phi_{lm}(t)$.
We identify a single orbit using two consecutive minima (maxima) of the strain amplitude $A_{lm}(t)$ corresponding to  apocenter (pericenter) passages. We compute the Fourier phase $\Psi$ and interpolate it with splines (which are differentiable) to avoid propagating finite-difference numerical errors to $\dot{\Psi}$ and~$\ddot{\Psi}$. %

\begin{figure} 
	\centering
	\subfloat{\includegraphics[width=\columnwidth]{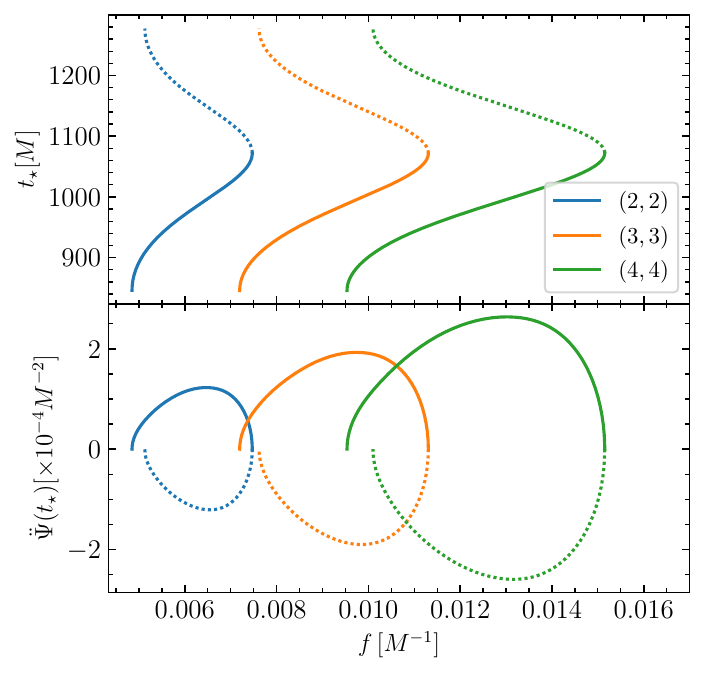}} %
	\caption{Stationary point $t_\star$ (top panel) and second derivative of the Fourier phase $\ddot{\Psi}$ (bottom panel) as a function of the frequency $f$ for a single orbit of the eccentric NR simulation SXS:BBH:1372. %
The halves of the orbit are indicated with solid and dotted curves. Colors indicate the different harmonics, with $(l,m)=$ (2,2) in blue, (3,3) in orange, and (4,4) in green.}
	\label{fig:three}
\end{figure}

Figure~\ref{fig:three} shows the stationary points $t_{\star}$ and the second derivative $\ddot{\Psi}$ as a function of the frequency~$f$ for each harmonic. 
The evolution of $\ddot{\Psi}$ describes a loop, that, however, does not close because of the %
radiation reaction. %
For each half orbit, there exists a finite region $f^-< f< f^+$  
where the phase $\Psi$ has a stationary point, while this cannot be found outside that range. The limiting case $f=f^\pm$, 
corresponding to a saddle point in $\Psi$, is a catastrophe, marking the appearance/disappearance of the stationary point itself. This is one of our key results: {NR waveforms are subjected to catastrophes}.

\subsection{Weighted average}
\label{sec:four}

\begin{figure} 
	\centering
	\subfloat{\includegraphics[width=\columnwidth]{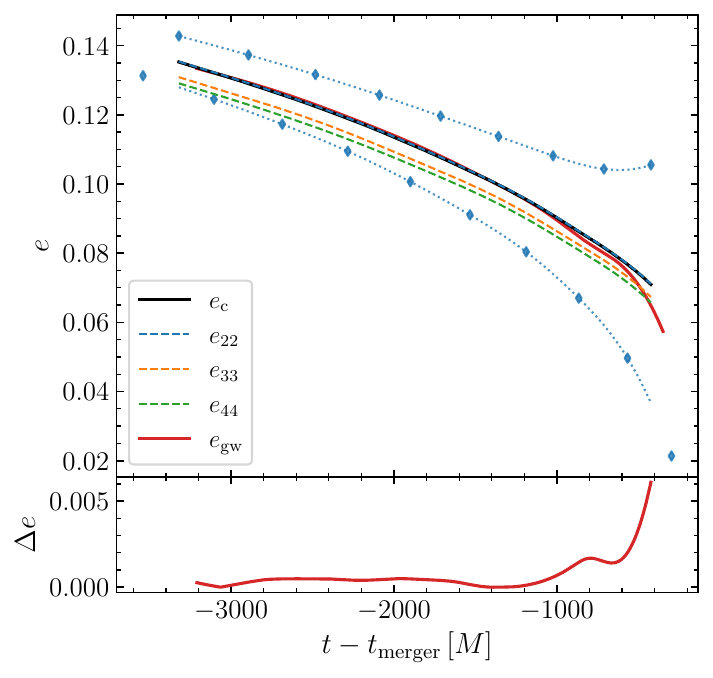}} 
	\caption{Eccentricity evolution in the  NR simulation SXS:BBH:1372. The blue dotted curves in the top panel show the upper (lower) bound on eccentricity estimated using half orbits from the apocenter (pericenter) to the pericenter (apocenter) of the (2,2) harmonic. The mean of these two curves is shown in dashed blue and corresponds to our eccentricity estimate for the (2,2) mode. The mean evolutions of the eccentricity computed from the (3,3) and (4,4) harmonics are shown in dashed orange and green, respectively. For each mode, the first and the last points are excluded to avoid extrapolation errors. The weighted average \ecat is shown in solid black and almost coincides with the $e_{22}$ estimate. The \egw estimate is shown in red. For this simulation, the reference eccentricity reported in the SXS catalog is $e\simeq0.21$ at $t - t_{\rm merger} = -3870 M$. The lower panel shows the absolute residuals between \ecat and~\egw.}%
	\label{fig:four}
\end{figure}

Our eccentricity estimate for the simulation SXS:BBH:1372 is shown in Fig.~\ref{fig:four}. We evaluate Eqs.~(\ref{eq:e22})--(\ref{eq:Phi33}), numerically  solve Eq.~(\ref{eq:r_44}) for each half orbit, and assign to each eccentricity estimate the average time between the two orbital extrema. This procedure ensures the Newtonian limit is respected.

Each harmonic provides a slightly different value of the eccentricity. We suspect this is because the mapping between frequency and eccentricity derived in Sec.~\ref{sec:two} makes use of the leading PN order for each mode---we leave the generalization to higher orders for future work. 
But these estimates should not be treated equally. They all come from the same waveform where each harmonic contributes differently. This intuition can be easily implemented with a weighted average, such that the dominant (subdominant) emission mode contributes more (less) to the final eccentricity estimate. More specifically, we consider the energy flux of each harmonic
$	\dot{E}_{lm}\propto  \left|\dd {h}_{lm} / \dd t\right |^2$ \cite{2008GReGr..40.1705R}
and weight each eccentricity contribution accordingly. Our final estimate is
\begin{equation}
\ecat = \frac{\displaystyle \sum_{l,m} e_{l m} \left|\frac{\dd {h}_{lm}} { \dd t}\right |^2 }{\displaystyle \sum_{l,m} \left|\frac{\dd {h}_{lm}} { \dd t}\right |^2 } \,,
\label{weave}
\end{equation}
where the sums are taken over all the modes for which a mapping is available, in this paper $(l,m) = (2,2), (3,3), and (4,4)$.

For the simulation shown in Fig.~\ref{fig:four}, we find the eccentricity \ecat evolves from $\sim 0.14$ to $\sim 0.08$. These values are not compatible with the PN-based eccentricity reported in the metadata of the SXS catalog, which is $\sim0.21$ at reference time $t - t_{\rm merger} = -3870 M$. 
The normalized weight associated with the (2,2) mode at $t-t_{\rm merger} = -3320 M$ is $\sim 97.8\%$, for the (3,3) mode is  $\sim 2.1\%$, while the (4,4) mode is responsible for only $0.01\%$ of the energy flux.%
This explains why \ecat in Fig.~\ref{fig:four} essentially coincides with $e_{22}$. %

Our estimate \ecat is in good agreement with the value estimate of Ref.~\cite{2023PhRvD.108j4007S}, hereafter \egw, which we recomputed for this simulation using their code. Residuals $\Delta e$ are of $\mathcal{O}\!\left(10^{-3}\right)$. This is not surprising because, as explored in Sec.~\ref{compegw},  $\egw$ is equivalent to our estimator when restricted to the dominant (2,2) mode.
In Fig.~\ref{fig:four}, our eccentricity evolution as a function of time is smoother than that of Ref.~\cite{2023PhRvD.108j4007S}, which instead undergoes small oscillations. We believe that is a spurious feature due to different implementations and numerical errors. %

\subsection{Numerical relativity catalogs}
\label{sec:five}

\begin{figure*} 
	\centering
{\includegraphics[width=\textwidth]{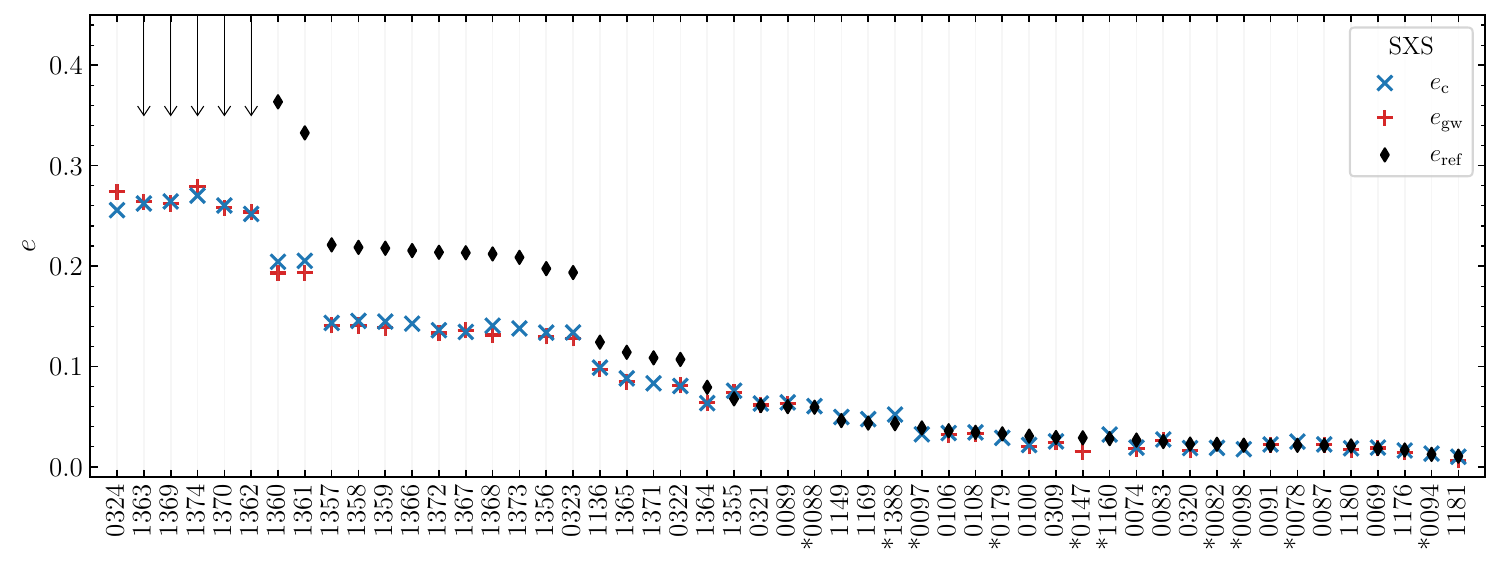}} \\[5pt]{\includegraphics[width=\textwidth]{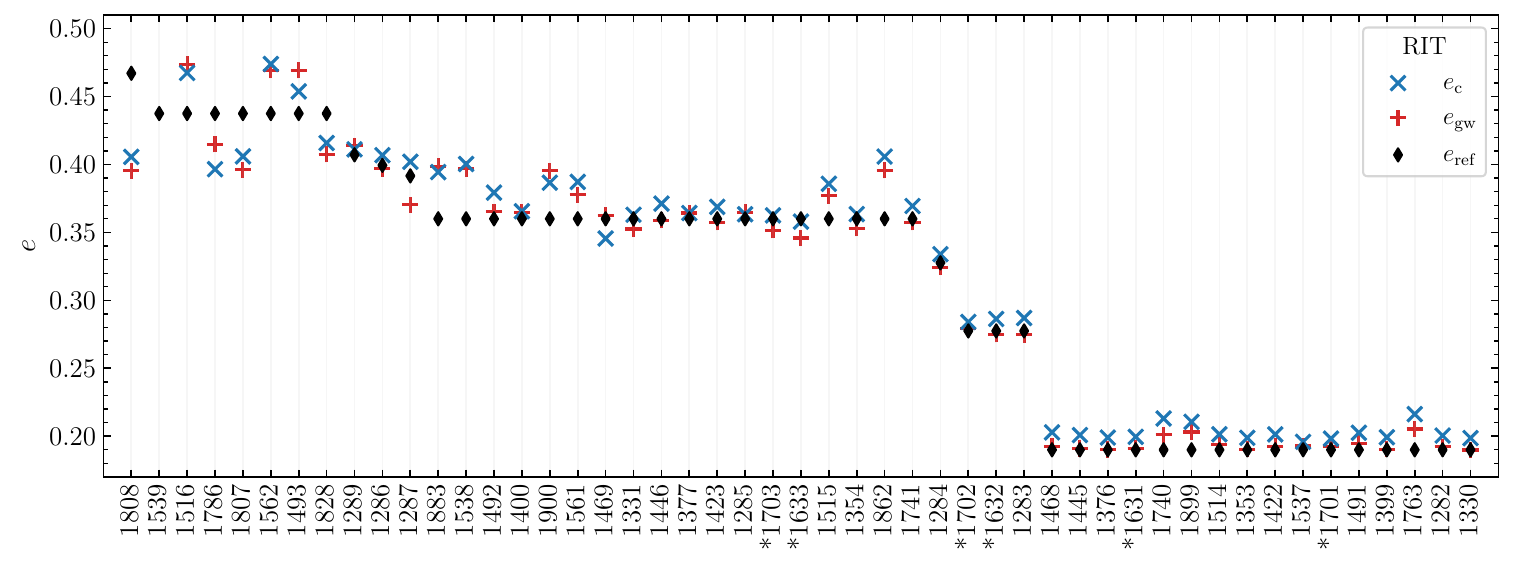}} 
	\caption{%
	Eccentric NR catalogs. The top (bottom) panel shows eccentricity estimates for the most eccentric simulations in the SXS \cite{2019CQGra..36s5006B}  (RIT \cite{2022PhRvD.105l4010H}) catalog.  %
	Black diamonds indicate the nominal eccentricity value reported in the catalog at the reference time. In a few cases, this estimate is provided as an upper limit (down-pointing arrows), and in a few others it is not provided at all. Blue crosses refer to our \ecat estimates measured at the earliest possible time past junk radiation. Similarly, red markers indicate \egw from Ref.~\cite{2023PhRvD.108j4007S}. Stars next to the ID numbers on the $x$ axis identify simulations with precessing effective spin  $\chi_{\rm p} > 0.01$. }
	\label{fig:five}
\end{figure*}

We apply our findings to 100 NR simulations from the SXS \cite{2019CQGra..36s5006B} and RIT \cite{2022PhRvD.105l4010H} groups. In particular, we select simulations with reported eccentricity $e\gtrsim 0.01$ lasting at least seven orbits. The results of our analysis are  presented in Fig.~\ref{fig:five} using the first available estimate past junk radiation. %
Additional details are provided in  Appendix~\ref{app:A}.

In the case of the SXS simulations, we observe discrepancies of up to 15\% at eccentricities $\gtrsim 0.1$ when comparing \ecat against the values reported in the catalog. Their eccentricities are estimated by fitting the orbital-frequency time derivative over the first $\ssim 2$ orbits with a PN ansatz. %
In particular, we find that the majority of their values overestimate the eccentricity compared to \ecat, suggesting that even the most eccentric simulations in the SXS catalog are closer to quasicircular than what was previously reported.

 For the RIT simulations, their nominal eccentricities are reported assuming $e \propto r^2 \dd^2 r/ \dd t^2$,  where $r$ is the coordinate separation.
 We find the values indicated in the RIT catalog are in better agreement with \ecat, with differences of a few \% and up to 10\% for eccentricities $\gtrsim 0.36$. 

We successfully measured the eccentricity for simulations that either do not present a catalog estimate (SXS:BBH:0324) or that only present an upper limit (SXS:BBH:1362, 1363, 1369, 1370, and 1374). For a particular case (SXS:BBH:0147), the combination of significant spin precession and too-short signal makes our current implementation inconclusive because amplitude peaks cannot be confidently identified. 
We can compute eccentricities in several cases for which the \egw code cannot provide an estimate (SXS:BBH:0078, 0082, 0088, 0094, 0097, 0098, 0179, 1149, 1160, 1169, 1366, 1371, 1373, and 1388). This is due to code implementation, which returns an error for highly precessing systems, and not to the definition itself. %
For the RIT runs, the simulations with the largest discrepancies (RIT:eBBH:1539, 1808, and 1862) correspond to the shorter waveforms. Shorter waveforms correspond to more rapidly evolving binaries that are closer to merger and it is more difficult to extract eccentricity from these. All methods struggle in this regime.

There is an important caveat in this comparison, which is related to the time, or frequency, at which the eccentricity is measured. NR catalogs report the value of $e$ at some reference time past junk radiation, and this depends on the length of the simulations themselves. Estimating \ecat  requires at least one pericenter and one apocenter passage (which are identified from the waveform), resulting in an eccentricity measure at some marginally larger time. This time difference mostly affects the analysis of highly eccentric and short waveforms, where the eccentricity evolves faster. This partially, but not fully, mitigates the reported differences. 
The times of eccentricity measurements are reported in Appendix \ref{app:A}.

While the mapping derived in Sec.~\ref{sec:two} is based on nonspinning PN waveforms, one can still estimate \ecat even for precessing binaries whenever the eccentricity modulation can be confidently identified in the waveform. This is because the orbital timescale on which one measures the eccentricity is typically smaller than the precession timescale. However, such timescale separation breaks down close to merger, and the two effects become comparable \cite{2023MNRAS.519.5352R}. With the procedure presented here, we can confidently estimate the eccentricity for the longer precessing waveforms in the catalogs (e.g. SXS:BBH:1388 or RIT:eBBH:1702) but struggle with shorter ones (e.g. SXS:BH:0147).

\subsection{Comparison with \egw}
\label{compegw}

Overall, we find that the eccentricity \ecat we extract from NR waveforms disagrees with the nominal values reported in the catalogs but largely agrees with \egw from Refs.~\cite{2023PhRvD.108j4007S,2022PhRvD.106l4040R}, 
which we report here for completeness:
 \begin{align}
 	e_{\rm gw} =& \cos\left(\psi/3\right) - \sqrt{3}\sin\left(\psi/3\right) \,\, , \\
	\psi =& \arctan\left(\frac{1-e^2_{\omega_{22}}}{2e_{\omega_{22}}}\right)\,\, , \\
	e_{\omega_{22}} =& \frac{\sqrt{\omega^p_{22}(t)} - \sqrt{\omega^a_{22}(t)}}{\sqrt{\omega^p_{22}(t)} + \sqrt{\omega^a_{22}(t)}}\,\, .
 \end{align} 
 Indeed, some nontrivial algebraic manipulation shows that the (2,2)-mode estimate presented in Eqs.~(\ref{eq:e22})--(\ref{eq:theta22}) is formally equivalent to the expression above.
The median of the absolute difference between the two definitions across the two sets of simulations is $\ssim 6.5 \times 10^{-3}$, and we report that they differ by more than $1\%$ in $\sim8\%$ ($\sim39\%$) of the cases among the SXS (RIT) simulations considered in this analysis. The larger discrepancies occurring in the second set are mostly due to  different measuring times (see Appendix~\ref{app:A}). %

We stress that the derivations of \ecat and \egw are different, with the former relying on the breakdown of the SPA and the latter relying on the instantaneous GW phase. In particular, the construction of \egw requires adjustments to ensure the resulting eccentricity estimator has the correct Newtonian limit. On the other hand, this is automatically implemented in our procedure, which makes explicit use of the said limit to derive the mapping between the catastrophe frequencies and the eccentricity. Our estimator \ecat includes emission modes of higher order, unlike $\egw$, which was explicitly restricted to the dominant harmonic. That said, the (2,2) mode is the most important contribution entering the weighted average of Eq.~(\ref{weave}).

\section{Conclusions} 

\label{sec:six}
Catastrophe theory provides an ideal framework to define the orbital eccentricity in GR. Measuring \ecat relies solely on the gravitational waveform and is thus fully gauge invariant. The Newtonian limit is also naturally built into the definition itself.

The eccentricity definition we put forward is fully generic---the only limitation is the resolution of the underlying NR simulations, as one needs to confidently identify peaks in the waveform amplitude. At present, this limits the applicability of our analysis to values of $\ecat \gtrsim 0.01$ (but note that eccentricities smaller than this threshold are unlikely to be distinguished from zero with current GW detectors~\cite{2018PhRvD..98h3028L}). We stress that this limitation is related only to the implementation and to the extremization routine, not to the theoretical foundations of this technique. On the contrary, at very high eccentricity ($e\gtrsim 0.6$),  the orbit-averaged approach \cite{1964PhRv..136.1224P} entering the mapping via Eqs.%
~(\ref{eq:dadt}) and (\ref{eq:dedt}) is no longer valid \cite{2019CQGra..36r5003M, 2023PhRvD.108l4055F}. %
This is due to the derivation strategy pursued in this paper and does not limit the applicability of catastrophe theory to eccentric BH binaries. %

In this paper, the $f_\pm \to e$ mapping was modeled using the $(l,m)=(2,2)$, $(3,3)$, and $(4,4)$ harmonics at their leading PN order. Our approach is generic and can be generalized to other modes using the appropriate PN expressions.  %
This will be increasingly relevant for large values of the eccentricity because power leaks into higher and  higher harmonics~\cite{1963PhRv..131..435P}. Future work will explore the suitability of catastrophe theory to define the second parameter describing eccentric systems, i.e. one of the orbital anomalies. %

Extracting $\ecat$ only requires the waveform time series and can thus be applied to GW signals detected with current and future interferometers. In the short term, we envision this as a postprocessing exercise on high-level GW data, namely, posterior samples. As such, using \ecat does not require changing the eccentricity definition internally employed by waveform approximants. This contributes to current attempts aimed at properly comparing the notions of eccentricity used in astrophysical models and those used in GW data analysis \cite{2023PhRvD.107f4024B,2024PhRvD.110j4002B,2024ApJ...969..132V}. %
In the long run, \ecat estimates might be exploited when building new waveform approximants, thus contributing to making models closer to the physical, gauge-invariant observables.

\acknowledgments
We thank Md Arif Shaikh, 
Harald Pfeiffer, Vijay Varma, Antoni Ramos Buades, Enrico Barausse, Costantino Pacilio, and Bruno Giacomazzo for discussions.
 M.B., N.L, D.G., and G.F are supported by ERC Starting Grant No.~945155--GWmining, 
Cariplo Foundation Grant No.~2021-0555, MUR PRIN Grant No.~2022-Z9X4XS, 
MUR Grant “Progetto Dipartimenti di Eccellenza 2023--2027” (BiCoQ),
and the ICSC National Research Centre funded by NextGenerationEU.  
D.G. is supported by MSCA Fellowships No.~101064542--StochRewind and  No.~101149270--ProtoBH.
Computational work was performed at CINECA with allocations 
through INFN and Bicocca.

\newpage

\bibliography{eccentriccatastrophes}

\onecolumngrid

\clearpage

\appendix

\section{Properties of the NR simulations}
\label{app:A}

We provide details on the NR simulations analyzed in this paper. Table~\ref{bigtable}  reports the mass ratio $q$, the effective spin $\chi_{\rm eff}$, the precessing parameter $\chi_{\rm p}$, and the number of orbits $n_{\rm orb}$. For each simulation, we list the nominal eccentricity $e_{\rm ref}$ provided in the catalog metadata, which is measured at time $t_{\rm ref}$. The estimates $e_{22}$, $e_{33}$, and $e_{44}$ are evaluated at the earliest possible times past junk radiation.The \ecat values refer to the first common time at which the eccentricities for the three harmonics are defined $t_{\rm c}$. For \egw, we use the code of Ref.~\cite{2023PhRvD.108j4007S}, selecting their ``amplitude'' method flag, and the estimates \egw are measured at the earliest time $t_{\rm gw}$, as defined in their code.

\renewcommand*{\arraystretch}{1.1}

\begin{longtable}{c@{\hspace{0.1cm}}|@{\hspace{0.1cm}}ccccccc@{\hspace{0.1cm}}|@{\hspace{0.1cm}}ccc@{\hspace{0.1cm}}|@{\hspace{0.1cm}}ccccccccc@{\hspace{0.1cm}}|@{\hspace{0.1cm}}ccc}
	& $q$ & & $\chi_{\rm eff}$ & & $\chi_{\rm p}$ & & $n_{\rm orb}$ & $e_{\rm ref}$ & & $t_{\rm ref} [M]$ & $e_{22}$  & & $e_{33}$ & & $e_{44}$ & & $e_{\rm c}$ & & $t_{\rm c} [M]$ & $e_{\rm gw}$ & & $t_{\rm gw} [M]$ \\[3pt]
	\hline \hline
	&&&&&&&&&&&&&&&&&&&&&&\\[-8pt]
	\endfirsthead

	 & $q$ & & $\chi_{\rm eff}$ & & $\chi_{\rm p}$ & & $n_{\rm orb}$ & $e_{\rm ref}$ & & $t_{\rm ref} [M]$ & $e_{22}$  & & $e_{33}$ & & $e_{44}$ & & $e_{\rm c}$ & & $t_{\rm c} [M]$ & $e_{\rm gw}$ & & $t_{\rm gw} [M]$ \\[3pt]
	\hline \hline 
	&&&&&&&&&&&&&&&&&&&&&&\\[-8pt]
	\endhead
	
        SXS:BBH:0324 & $0.82$ & & $-1.7\!\times\! 10^{-2}$ & & $2.1\!\times\! 10^{-8}$ & & $13.04$ & --- & & $-2790$ & $0.256$ & & $0.260$ & & $0.301$ & & $0.256$ & & $-2148$ & $0.274$ & & $-2052$ \\
        SXS:BBH:1363 & $1.00$ & & $3.5\!\times\! 10^{-6}$ & & $5.3\!\times\! 10^{-8}$ & & $12.18$ & $<1.80$ & & $-2509$ & $0.262$ & & --- & & $0.292$ & & $0.262$ & & $-1884$ & $0.264$ & & $-1773$ \\
        SXS:BBH:1369 & $0.50$ & & $4.4\!\times\! 10^{-5}$ & & $5.1\!\times\! 10^{-8}$ & & $13.92$ & $<1.80$ & & $-2895$ & $0.265$ & & $0.281$ & & $0.329$ & & $0.264$ & & $-2346$ & $0.262$ & & $-2254$ \\
        SXS:BBH:1374 & $0.33$ & & $-6.5\!\times\! 10^{-5}$ & & $1.3\!\times\! 10^{-7}$ & & $15.59$ & $<1.70$ & & $-3413$ & $0.270$ & & $0.277$ & & $0.275$ & & $0.270$ & & $-2975$ & $0.279$ & & $-2644$ \\
        SXS:BBH:1370 & $0.50$ & & $2.8\!\times\! 10^{-5}$ & & $9.2\!\times\! 10^{-8}$ & & $13.23$ & $<1.70$ & & $-2649$ & $0.260$ & & $0.266$ & & $0.269$ & & $0.260$ & & $-2148$ & $0.258$ & & $-2038$ \\
        SXS:BBH:1362 & $1.00$ & & $1.2\!\times\! 10^{-5}$ & & $2.0\!\times\! 10^{-8}$ & & $12.28$ & $<1.70$ & & $-2567$ & $0.252$ & & --- & & $0.422$ & & $0.252$ & & $-1709$ & $0.254$ & & $-1812$ \\
        SXS:BBH:1360 & $1.00$ & & $-4.5\!\times\! 10^{-7}$ & & $7.0\!\times\! 10^{-9}$ & & $13.14$ & $0.364$ & & $-2791$ & $0.204$ & & --- & & $0.199$ & & $0.204$ & & $-2363$ & $0.193$ & & $-2043$ \\
        SXS:BBH:1361 & $1.00$ & & $-6.5\!\times\! 10^{-6}$ & & $4.2\!\times\! 10^{-8}$ & & $12.98$ & $0.333$ & & $-2713$ & $0.205$ & & --- & & $0.204$ & & $0.205$ & & $-2316$ & $0.193$ & & $-2003$ \\
        SXS:BBH:1357 & $1.00$ & & $-1.9\!\times\! 10^{-5}$ & & $2.3\!\times\! 10^{-8}$ & & $14.76$ & $0.221$ & & $-3284$ & $0.143$ & & --- & & $0.138$ & & $0.143$ & & $-2647$ & $0.141$ & & $-2537$ \\
        SXS:BBH:1358 & $1.00$ & & $-2.6\!\times\! 10^{-5}$ & & $6.6\!\times\! 10^{-8}$ & & $14.08$ & $0.219$ & & $-3056$ & $0.145$ & & --- & & $0.140$ & & $0.145$ & & $-2643$ & $0.141$ & & $-2324$ \\
        SXS:BBH:1359 & $1.00$ & & $-1.5\!\times\! 10^{-5}$ & & $3.7\!\times\! 10^{-9}$ & & $13.75$ & $0.218$ & & $-2948$ & $0.145$ & & --- & & $0.139$ & & $0.145$ & & $-2527$ & $0.139$ & & $-2211$ \\
        SXS:BBH:1366 & $0.50$ & & $1.2\!\times\! 10^{-4}$ & & $3.1\!\times\! 10^{-6}$ & & $15.62$ & $0.215$ & & $-3477$ & $0.143$ & & $0.144$ & & $0.136$ & & $0.143$ & & $-3053$ & --- & & --- \\
        SXS:BBH:1372 & $0.33$ & & $3.1\!\times\! 10^{-5}$ & & $2.4\!\times\! 10^{-7}$ & & $17.66$ & $0.214$ & & $-3870$ & $0.136$ & & $0.137$ & & $0.130$ & & $0.136$ & & $-3320$ & $0.133$ & & $-3210$ \\
        SXS:BBH:1367 & $0.50$ & & $-4.0\!\times\! 10^{-5}$ & & $9.1\!\times\! 10^{-8}$ & & $15.31$ & $0.213$ & & $-3027$ & $0.134$ & & $0.135$ & & $0.128$ & & $0.134$ & & $-2513$ & $0.136$ & & $-2618$ \\
        SXS:BBH:1368 & $0.50$ & & $-7.7\!\times\! 10^{-5}$ & & $9.9\!\times\! 10^{-8}$ & & $15.03$ & $0.212$ & & $-3259$ & $0.141$ & & $0.142$ & & $0.135$ & & $0.141$ & & $-2842$ & $0.131$ & & $-2522$ \\
        SXS:BBH:1373 & $0.33$ & & $2.1\!\times\! 10^{-5}$ & & $9.1\!\times\! 10^{-8}$ & & $17.31$ & $0.209$ & & $-3785$ & $0.138$ & & $0.139$ & & $0.132$ & & $0.138$ & & $-3432$ & --- & & --- \\
        SXS:BBH:1356 & $1.00$ & & $2.4\!\times\! 10^{-6}$ & & $2.9\!\times\! 10^{-8}$ & & $22.34$ & $0.197$ & & $-6448$ & $0.134$ & & --- & & $0.129$ & & $0.134$ & & $-5862$ & $0.130$ & & $-5455$ \\
        SXS:BBH:0323 & $0.82$ & & $-1.7\!\times\! 10^{-2}$ & & $7.6\!\times\! 10^{-8}$ & & $14.63$ & $0.194$ & & $-3212$ & $0.134$ & & $0.135$ & & $0.128$ & & $0.134$ & & $-2806$ & $0.128$ & & $-2485$ \\
        SXS:BBH:1136 & $1.00$ & & $-7.5\!\times\! 10^{-1}$ & & $5.1\!\times\! 10^{-8}$ & & $9.51$ & $0.124$ & & $-1820$ & $0.099$ & & --- & & $0.093$ & & $0.099$ & & $-1453$ & $0.097$ & & $-1352$ \\
        SXS:BBH:1365 & $0.50$ & & $2.3\!\times\! 10^{-5}$ & & $4.1\!\times\! 10^{-7}$ & & $16.06$ & $0.114$ & & $-3554$ & $0.088$ & & $0.088$ & & $0.085$ & & $0.088$ & & $-3180$ & $0.085$ & & $-2853$ \\
        SXS:BBH:1371 & $0.33$ & & $4.1\!\times\! 10^{-5}$ & & $2.2\!\times\! 10^{-7}$ & & $18.17$ & $0.109$ & & $-4102$ & $0.083$ & & $0.083$ & & $0.080$ & & $0.083$ & & $-3699$ & --- & & --- \\
        SXS:BBH:0322 & $0.82$ & & $-1.7\!\times\! 10^{-2}$ & & $4.6\!\times\! 10^{-8}$ & & $14.97$ & $0.107$ & & $-2990$ & $0.081$ & & $0.080$ & & $0.077$ & & $0.081$ & & $-2499$ & $0.081$ & & $-2604$ \\
        SXS:BBH:1364 & $0.50$ & & $2.0\!\times\! 10^{-5}$ & & $2.8\!\times\! 10^{-7}$ & & $16.14$ & $0.079$ & & $-3272$ & $0.063$ & & $0.063$ & & $0.061$ & & $0.063$ & & $-2785$ & $0.064$ & & $-2889$ \\
        SXS:BBH:1355 & $1.00$ & & $-1.6\!\times\! 10^{-5}$ & & $2.5\!\times\! 10^{-8}$ & & $13.90$ & $0.068$ & & $-2936$ & $0.076$ & & --- & & $0.072$ & & $0.076$ & & $-2452$ & $0.074$ & & $-2347$ \\
        SXS:BBH:0321 & $0.82$ & & $-1.7\!\times\! 10^{-2}$ & & $1.4\!\times\! 10^{-7}$ & & $14.98$ & $0.061$ & & $-3305$ & $0.063$ & & $0.062$ & & $0.060$ & & $0.063$ & & $-2761$ & $0.062$ & & $-2653$ \\
        SXS:BBH:0089 & $1.00$ & & $-2.5\!\times\! 10^{-1}$ & & $2.3\!\times\! 10^{-10}$ & & $31.06$ & $0.060$ & & $-11363$ & $0.064$ & & --- & & $0.063$ & & $0.064$ & & $-10662$ & $0.063$ & & $-10154$ \\
        SXS:BBH:0088 & $1.00$ & & $1.2\!\times\! 10^{-5}$ & & $5.0\!\times\! 10^{-1}$ & & $31.87$ & $0.059$ & & $-11285$ & $0.061$ & & --- & & $0.060$ & & $0.061$ & & $-10560$ & --- & & --- \\
        SXS:BBH:1149 & $0.33$ & & $6.8\!\times\! 10^{-1}$ & & $7.7\!\times\! 10^{-7}$ & & $24.10$ & $0.046$ & & $-5169$ & $0.050$ & & $0.049$ & & $0.048$ & & $0.050$ & & $-4753$ & --- & & --- \\
        SXS:BBH:1169 & $0.33$ & & $-6.7\!\times\! 10^{-1}$ & & $2.7\!\times\! 10^{-6}$ & & $22.07$ & $0.044$ & & $-6426$ & $0.048$ & & $0.047$ & & $0.046$ & & $0.048$ & & $-5729$ & --- & & --- \\
        SXS:BBH:1388 & $0.33$ & & $4.2\!\times\! 10^{-1}$ & & $5.8\!\times\! 10^{-1}$ & & $23.60$ & $0.043$ & & $-5330$ & $0.052$ & & $0.046$ & & $0.047$ & & $0.052$ & & $-4942$ & --- & & --- \\
        SXS:BBH:0097 & $0.67$ & & $1.3\!\times\! 10^{-4}$ & & $5.0\!\times\! 10^{-1}$ & & $19.24$ & $0.039$ & & $-5074$ & $0.032$ & & $0.039$ & & $0.033$ & & $0.032$ & & $-4386$ & --- & & --- \\
        SXS:BBH:0106 & $0.20$ & & $1.9\!\times\! 10^{-7}$ & & $4.9\!\times\! 10^{-10}$ & & $21.32$ & $0.036$ & & $-4669$ & $0.034$ & & $0.033$ & & $0.032$ & & $0.034$ & & $-4306$ & $0.032$ & & $-3980$ \\
        SXS:BBH:0108 & $0.20$ & & $-4.2\!\times\! 10^{-1}$ & & $6.2\!\times\! 10^{-11}$ & & $20.73$ & $0.034$ & & $-5226$ & $0.034$ & & $0.034$ & & $0.033$ & & $0.034$ & & $-4703$ & $0.033$ & & $-4333$ \\
        SXS:BBH:0179 & $0.67$ & & $6.5\!\times\! 10^{-1}$ & & $8.7\!\times\! 10^{-2}$ & & $23.75$ & $0.033$ & & $-5714$ & $0.029$ & & $0.030$ & & $0.028$ & & $0.029$ & & $-5368$ & --- & & --- \\
        SXS:BBH:0100 & $0.67$ & & $1.2\!\times\! 10^{-7}$ & & $7.3\!\times\! 10^{-10}$ & & $27.36$ & $0.031$ & & $-8472$ & $0.022$ & & $0.021$ & & $0.021$ & & $0.022$ & & $-7969$ & $0.021$ & & $-7526$ \\
        SXS:BBH:0309 & $0.82$ & & $-1.6\!\times\! 10^{-2}$ & & $1.2\!\times\! 10^{-7}$ & & $15.75$ & $0.029$ & & $-3258$ & $0.025$ & & $0.024$ & & $0.024$ & & $0.025$ & & $-2806$ & $0.025$ & & $-2906$ \\
        SXS:BBH:0147 & $1.00$ & & $-1.1\!\times\! 10^{-3}$ & & $5.0\!\times\! 10^{-1}$ & & $7.12$ & $0.029$ & & $-882$ & --- & & --- & & --- & & --- & & --- & $0.015$ & & $-510$ \\
        SXS:BBH:1160 & $0.33$ & & $2.2\!\times\! 10^{-1}$ & & $5.9\!\times\! 10^{-1}$ & & $22.36$ & $0.028$ & & $-4901$ & $0.032$ & & $0.033$ & & $0.032$ & & $0.032$ & & $-4489$ & --- & & --- \\
        SXS:BBH:0074 & $1.00$ & & $1.3\!\times\! 10^{-7}$ & & $1.2\!\times\! 10^{-9}$ & & $26.23$ & $0.027$ & & $-8037$ & $0.019$ & & --- & & $0.018$ & & $0.019$ & & $-7548$ & $0.018$ & & $-7105$ \\
        SXS:BBH:0083 & $1.00$ & & $2.5\!\times\! 10^{-1}$ & & $2.8\!\times\! 10^{-9}$ & & $32.40$ & $0.025$ & & $-11067$ & $0.027$ & & --- & & $0.027$ & & $0.027$ & & $-10491$ & $0.027$ & & $-9715$ \\
        SXS:BBH:0320 & $0.82$ & & $-1.7\!\times\! 10^{-2}$ & & $1.0\!\times\! 10^{-7}$ & & $13.49$ & $0.023$ & & $-2447$ & $0.019$ & & $0.018$ & & $0.018$ & & $0.019$ & & $-1981$ & $0.017$ & & $-2067$ \\
        SXS:BBH:0082 & $0.67$ & & $7.8\!\times\! 10^{-4}$ & & $5.0\!\times\! 10^{-1}$ & & $19.77$ & $0.023$ & & $-5310$ & $0.019$ & & $0.025$ & & $0.018$ & & $0.019$ & & $-4588$ & --- & & --- \\
        SXS:BBH:0098 & $0.67$ & & $6.9\!\times\! 10^{-4}$ & & $5.0\!\times\! 10^{-1}$ & & $27.53$ & $0.022$ & & $-8614$ & $0.018$ & & $0.025$ & & $0.017$ & & $0.018$ & & $-8032$ & --- & & --- \\
        SXS:BBH:0091 & $1.00$ & & $9.8\!\times\! 10^{-8}$ & & $2.2\!\times\! 10^{-9}$ & & $34.18$ & $0.022$ & & $-12498$ & $0.022$ & & --- & & $0.022$ & & $0.022$ & & $-11647$ & $0.022$ & & $-11154$ \\
        SXS:BBH:0078 & $0.67$ & & $7.9\!\times\! 10^{-4}$ & & $5.0\!\times\! 10^{-1}$ & & $23.21$ & $0.022$ & & $-6811$ & $0.025$ & & $0.016$ & & $0.024$ & & $0.025$ & & $-6281$ & --- & & --- \\
        SXS:BBH:0087 & $1.00$ & & $1.2\!\times\! 10^{-7}$ & & $1.4\!\times\! 10^{-9}$ & & $29.96$ & $0.021$ & & $-10161$ & $0.022$ & & --- & & $0.023$ & & $0.022$ & & $-9387$ & $0.022$ & & $-8924$ \\
        SXS:BBH:1180 & $0.33$ & & $-1.1\!\times\! 10^{-4}$ & & $3.2\!\times\! 10^{-7}$ & & $15.34$ & $0.021$ & & $-2781$ & $0.019$ & & $0.018$ & & $0.018$ & & $0.019$ & & $-2389$ & $0.017$ & & $-2476$ \\
        SXS:BBH:0069 & $1.00$ & & $1.2\!\times\! 10^{-7}$ & & $1.6\!\times\! 10^{-9}$ & & $30.07$ & $0.018$ & & $-10245$ & $0.019$ & & --- & & $0.019$ & & $0.019$ & & $-9476$ & $0.019$ & & $-9010$ \\
        SXS:BBH:1176 & $0.33$ & & $-4.7\!\times\! 10^{-6}$ & & $1.3\!\times\! 10^{-7}$ & & $15.69$ & $0.017$ & & $-2900$ & $0.016$ & & $0.016$ & & $0.015$ & & $0.016$ & & $-2512$ & $0.015$ & & $-2594$ \\
        SXS:BBH:0094 & $0.67$ & & $3.5\!\times\! 10^{-4}$ & & $5.0\!\times\! 10^{-1}$ & & $21.42$ & $0.013$ & & $-5993$ & $0.014$ & & $0.018$ & & $0.015$ & & $0.013$ & & $-5369$ & --- & & --- \\
        SXS:BBH:1181 & $0.33$ & & $2.1\!\times\! 10^{-5}$ & & $9.5\!\times\! 10^{-8}$ & & $15.13$ & $0.011$ & & $-2884$ & $0.010$ & & $0.010$ & & $0.010$ & & $0.010$ & & $-2667$ & $0.007$ & & $-2339$ \\
\hline
        RIT:eBBH:1808 & $1.00$ & & $8.0\!\times\! 10^{-1}$ & & $0$ & & $11.75$ & $0.467$ & & $-2291$ & $0.406$ & & --- & & $0.440$ & & $0.406$ & & $-1363$ & $0.395$ & & $-1251$ \\
        RIT:eBBH:1539 & $0.17$ & & $3.8\!\times\! 10^{-6}$ & & $0$ & & $9.89$ & $0.438$ & & $-2154$ & $0.520$ & & $0.510$ & & $0.539$ & & $0.520$ & & $-1301$ & $0.517$ & & $-1412$ \\
        RIT:eBBH:1516 & $0.20$ & & $0$ & & $0$ & & $11.34$ & $0.438$ & & $-2436$ & $0.468$ & & $0.467$ & & $0.452$ & & $0.467$ & & $-1615$ & $0.474$ & & $-1724$ \\
        RIT:eBBH:1786 & $1.00$ & & $5.0\!\times\! 10^{-1}$ & & $0$ & & $11.37$ & $0.438$ & & $-2390$ & $0.398$ & & --- & & $0.395$ & & $0.397$ & & $-1391$ & $0.415$ & & $-1707$ \\
        RIT:eBBH:1807 & $1.00$ & & $8.0\!\times\! 10^{-1}$ & & $0$ & & $13.83$ & $0.438$ & & $-2924$ & $0.408$ & & --- & & $0.434$ & & $0.406$ & & $-2115$ & $0.396$ & & $-2023$ \\
        RIT:eBBH:1562 & $0.14$ & & $0$ & & $0$ & & $12.31$ & $0.438$ & & $-2621$ & $0.475$ & & $0.471$ & & $0.457$ & & $0.474$ & & $-1791$ & $0.469$ & & $-1688$ \\
        RIT:eBBH:1493 & $0.25$ & & $0$ & & $0$ & & $9.64$ & $0.438$ & & $-2044$ & $0.454$ & & $0.452$ & & $0.439$ & & $0.454$ & & $-1244$ & $0.469$ & & $-1688$ \\
        RIT:eBBH:1828 & $1.00$ & & $4.0\!\times\! 10^{-1}$ & & $0$ & & $10.54$ & $0.438$ & & $-2216$ & $0.418$ & & --- & & $0.403$ & & $0.416$ & & $-1434$ & $0.407$ & & $-1332$ \\
        RIT:eBBH:1289 & $1.00$ & & $0$ & & $0$ & & $9.36$ & $0.407$ & & $-2083$ & $0.411$ & & --- & & $0.413$ & & $0.411$ & & $-1273$ & $0.414$ & & $-1373$ \\
        RIT:eBBH:1286 & $1.00$ & & $0$ & & $0$ & & $10.07$ & $0.399$ & & $-2272$ & $0.407$ & & --- & & $0.408$ & & $0.407$ & & $-1448$ & $0.397$ & & $-1331$ \\
        RIT:eBBH:1287 & $1.00$ & & $0$ & & $0$ & & $10.49$ & $0.392$ & & $-2449$ & $0.402$ & & --- & & $0.397$ & & $0.402$ & & $-1615$ & $0.370$ & & $-1270$ \\
        RIT:eBBH:1883 & $0.50$ & & $-2.7\!\times\! 10^{-1}$ & & $0$ & & $8.96$ & $0.360$ & & $-2200$ & $0.394$ & & $0.391$ & & $0.390$ & & $0.394$ & & $-1307$ & $0.399$ & & $-1421$ \\
        RIT:eBBH:1538 & $0.17$ & & $0$ & & $0$ & & $12.94$ & $0.360$ & & $-3140$ & $0.400$ & & $0.405$ & & $0.385$ & & $0.400$ & & $-2005$ & $0.397$ & & $-2103$ \\
        RIT:eBBH:1492 & $0.25$ & & $0$ & & $0$ & & $18.31$ & $0.360$ & & $-4820$ & $0.379$ & & $0.378$ & & $0.378$ & & $0.379$ & & $-3657$ & $0.365$ & & $-3498$ \\
        RIT:eBBH:1400 & $0.60$ & & $0$ & & $0$ & & $13.91$ & $0.360$ & & $-3539$ & $0.366$ & & $0.365$ & & $0.354$ & & $0.366$ & & $-2403$ & $0.365$ & & $-2509$ \\
        RIT:eBBH:1900 & $1.00$ & & $-4.0\!\times\! 10^{-1}$ & & $0$ & & $9.69$ & $0.360$ & & $-2426$ & $0.388$ & & --- & & $0.398$ & & $0.387$ & & $-1530$ & $0.395$ & & $-1913$ \\
        RIT:eBBH:1561 & $0.14$ & & $0$ & & $0$ & & $23.84$ & $0.360$ & & $-6431$ & $0.387$ & & $0.392$ & & $0.389$ & & $0.387$ & & $-5252$ & $0.378$ & & $-5359$ \\
        RIT:eBBH:1469 & $0.33$ & & $0$ & & $0$ & & $16.09$ & $0.360$ & & $-4191$ & $0.376$ & & $0.373$ & & $0.368$ & & $0.345$ & & $-2233$ & $0.363$ & & $-2889$ \\
        RIT:eBBH:1331 & $0.90$ & & $0$ & & $0$ & & $13.35$ & $0.360$ & & $-3364$ & $0.363$ & & $0.362$ & & $0.351$ & & $0.363$ & & $-2233$ & $0.352$ & & $-2093$ \\
        RIT:eBBH:1446 & $0.40$ & & $0$ & & $0$ & & $15.42$ & $0.360$ & & $-3942$ & $0.371$ & & $0.369$ & & $0.364$ & & $0.371$ & & $-2796$ & $0.359$ & & $-2647$ \\
        RIT:eBBH:1377 & $0.70$ & & $0$ & & $0$ & & $13.38$ & $0.360$ & & $-3440$ & $0.364$ & & $0.363$ & & $0.354$ & & $0.364$ & & $-2306$ & $0.364$ & & $-2412$ \\
        RIT:eBBH:1423 & $0.50$ & & $0$ & & $0$ & & $14.46$ & $0.360$ & & $-3657$ & $0.369$ & & $0.367$ & & $0.360$ & & $0.369$ & & $-2519$ & $0.358$ & & $-2374$ \\
        RIT:eBBH:1285 & $1.00$ & & $0$ & & $0$ & & $13.02$ & $0.360$ & & $-3310$ & $0.364$ & & --- & & $0.357$ & & $0.363$ & & $-2172$ & $0.365$ & & $-2288$ \\
        RIT:eBBH:1703 & $1.00$ & & $0$ & & $7.0\!\times\! 10^{-1}$ & & $13.02$ & $0.360$ & & $-3324$ & $0.363$ & & --- & & $0.344$ & & $0.363$ & & $-2195$ & $0.352$ & & $-2063$ \\
        RIT:eBBH:1633 & $1.00$ & & $0$ & & $7.0\!\times\! 10^{-1}$ & & $13.40$ & $0.360$ & & $-3433$ & $0.358$ & & --- & & $0.344$ & & $0.358$ & & $-2296$ & $0.346$ & & $-2163$ \\
        RIT:eBBH:1515 & $0.20$ & & $2.0\!\times\! 10^{-6}$ & & $0$ & & $20.65$ & $0.360$ & & $-5480$ & $0.386$ & & $0.389$ & & $0.379$ & & $0.386$ & & $-4302$ & $0.377$ & & $-4409$ \\
        RIT:eBBH:1354 & $0.80$ & & $0$ & & $0$ & & $13.22$ & $0.360$ & & $-3388$ & $0.364$ & & $0.362$ & & $0.351$ & & $0.364$ & & $-2256$ & $0.353$ & & $-2116$ \\
        RIT:eBBH:1862 & $0.33$ & & $-2.0\!\times\! 10^{-1}$ & & $0$ & & $8.51$ & $0.360$ & & $-2151$ & $0.407$ & & $0.404$ & & $0.394$ & & $0.406$ & & $-1235$ & $0.396$ & & $-1133$ \\
        RIT:eBBH:1741 & $1.00$ & & $-5.0\!\times\! 10^{-1}$ & & $0$ & & $9.12$ & $0.360$ & & $-2229$ & $0.370$ & & --- & & $0.354$ & & $0.369$ & & $-1109$ & $0.358$ & & $-2374$ \\
        RIT:eBBH:1284 & $1.00$ & & $0$ & & $0$ & & $15.84$ & $0.328$ & & $-4372$ & $0.334$ & & --- & & $0.325$ & & $0.334$ & & $-3164$ & $0.324$ & & $-3020$ \\
        RIT:eBBH:1702 & $1.00$ & & $0$ & & $7.0\!\times\! 10^{-1}$ & & $20.98$ & $0.278$ & & $-6496$ & $0.284$ & & --- & & $0.290$ & & $0.284$ & & $-5183$ & $0.279$ & & $-5028$ \\
        RIT:eBBH:1632 & $1.00$ & & $0$ & & $7.0\!\times\! 10^{-1}$ & & $21.50$ & $0.278$ & & $-6666$ & $0.286$ & & --- & & $0.297$ & & $0.286$ & & $-5349$ & $0.275$ & & $-4928$ \\
        RIT:eBBH:1283 & $1.00$ & & $0$ & & $0$ & & $20.93$ & $0.278$ & & $-6466$ & $0.287$ & & --- & & $0.287$ & & $0.287$ & & $-5153$ & $0.275$ & & $-4732$ \\
        RIT:eBBH:1468 & $0.33$ & & $0$ & & $0$ & & $40.24$ & $0.190$ & & $-15055$ & $0.203$ & & $0.219$ & & $0.223$ & & $0.203$ & & $-13568$ & $0.193$ & & $-13696$ \\
        RIT:eBBH:1445 & $0.40$ & & $0$ & & $0$ & & $38.11$ & $0.190$ & & $-14182$ & $0.201$ & & $0.217$ & & $0.221$ & & $0.201$ & & $-12695$ & $0.191$ & & $-12823$ \\
        RIT:eBBH:1376 & $0.70$ & & $0$ & & $0$ & & $33.12$ & $0.190$ & & $-12234$ & $0.199$ & & $0.216$ & & $0.219$ & & $0.199$ & & $-10752$ & $0.190$ & & $-10554$ \\
        RIT:eBBH:1631 & $1.00$ & & $0$ & & $7.0\!\times\! 10^{-1}$ & & $32.31$ & $0.190$ & & $-11959$ & $0.200$ & & --- & & $0.231$ & & $0.200$ & & $-10481$ & $0.191$ & & $-10607$ \\
        RIT:eBBH:1740 & $1.00$ & & $-5.0\!\times\! 10^{-1}$ & & $0$ & & $26.82$ & $0.190$ & & $-10036$ & $0.213$ & & --- & & $0.226$ & & $0.213$ & & $-8540$ & $0.201$ & & $-8677$ \\
        RIT:eBBH:1899 & $1.00$ & & $-4.0\!\times\! 10^{-1}$ & & $0$ & & $28.30$ & $0.190$ & & $-10530$ & $0.211$ & & --- & & $0.222$ & & $0.211$ & & $-9033$ & $0.203$ & & $-8837$ \\
        RIT:eBBH:1514 & $0.20$ & & $0$ & & $0$ & & $43.30$ & $0.190$ & & $-15722$ & $0.201$ & & $0.217$ & & $0.223$ & & $0.201$ & & $-14227$ & $0.194$ & & $-14024$ \\
        RIT:eBBH:1353 & $0.80$ & & $0$ & & $0$ & & $32.46$ & $0.190$ & & $-12005$ & $0.199$ & & $0.216$ & & $0.220$ & & $0.199$ & & $-10524$ & $0.190$ & & $-10652$ \\
        RIT:eBBH:1422 & $0.50$ & & $0$ & & $0$ & & $35.11$ & $0.190$ & & $-13016$ & $0.201$ & & $0.218$ & & $0.220$ & & $0.201$ & & $-11534$ & $0.193$ & & $-11335$ \\
        RIT:eBBH:1537 & $0.17$ & & $0$ & & $0$ & & $20.90$ & $0.190$ & & $-6209$ & $0.195$ & & $0.207$ & & $0.192$ & & $0.196$ & & $-4795$ & $0.193$ & & $-4953$ \\
        RIT:eBBH:1701 & $1.00$ & & $0$ & & $7.0\!\times\! 10^{-1}$ & & $31.67$ & $0.190$ & & $-11702$ & $0.198$ & & --- & & $0.210$ & & $0.198$ & & $-10223$ & $0.192$ & & $-10356$ \\
        RIT:eBBH:1491 & $0.25$ & & $0$ & & $0$ & & $46.99$ & $0.190$ & & $-17776$ & $0.202$ & & $0.217$ & & $0.224$ & & $0.203$ & & $-16282$ & $0.195$ & & $-16079$ \\
        RIT:eBBH:1399 & $0.60$ & & $0$ & & $0$ & & $34.05$ & $0.190$ & & $-12651$ & $0.199$ & & $0.216$ & & $0.219$ & & $0.199$ & & $-11168$ & $0.190$ & & $-11296$ \\
        RIT:eBBH:1763 & $1.00$ & & $-8.0\!\times\! 10^{-1}$ & & $0$ & & $23.85$ & $0.190$ & & $-8996$ & $0.216$ & & --- & & $0.229$ & & $0.216$ & & $-7495$ & $0.205$ & & $-7287$ \\
        RIT:eBBH:1282 & $1.00$ & & $0$ & & $0$ & & $31.68$ & $0.190$ & & $-11706$ & $0.200$ & & --- & & $0.220$ & & $0.200$ & & $-10230$ & $0.192$ & & $-10357$ \\
        RIT:eBBH:1330 & $0.90$ & & $0$ & & $0$ & & $32.19$ & $0.190$ & & $-11894$ & $0.199$ & & $0.217$ & & $0.218$ & & $0.199$ & & $-10413$ & $0.190$ & & $-10215$\\
\multicolumn{5}{c}{$\;$}	\\[1pt]
\end{longtable}
\vspace{-1cm}
\begin{table}[h!]
	\caption{NR simulations analyzed in this paper. We refer to each run using the ID numbers reported in the SXS and RIT databases.%
	} %
	\label{bigtable}
\end{table}

\end{document}